\documentclass[10pt,twocolumn,superscriptaddress,english,10pt,prl,
english,10pt,prl,
floatfix,aps
]{revtex4-2}
\usepackage[T1]{fontenc}           					
\usepackage[utf8]{inputenc}
\usepackage{amsmath}
\usepackage{graphicx}
\usepackage{xspace}
\usepackage{physics}
\usepackage{verbatim}
\usepackage[backref=none,bookmarksnumbered=true,bookmarks=true,  	bookmarksopen=true,colorlinks=true,citecolor=blue,linkcolor=blue, 	anchorcolor=green,urlcolor=blue,unicode=false]{hyperref}

\makeatletter

\@ifundefined{textcolor}{}
{%
 \definecolor{BLACK}{gray}{0}
 \definecolor{WHITE}{gray}{1}
 \definecolor{RED}{rgb}{1,0,0}
 \definecolor{GREEN}{rgb}{0,1,0}
 \definecolor{BLUE}{rgb}{0,0,1}
 \definecolor{CYAN}{cmyk}{1,0,0,0}
 \definecolor{MAGENTA}{cmyk}{0,1,0,0}
 \definecolor{YELLOW}{cmyk}{0,0,1,0}
}

\usepackage{soul}
\usepackage{braket}
\usepackage{ulem}
\usepackage[backref=none,
bookmarksnumbered=true,
bookmarks=true,
bookmarksopen=true,
colorlinks=true,
citecolor=blue,
linkcolor=blue,
anchorcolor=green,
urlcolor=blue,unicode=false]{hyperref}

\let\baraccent=\=  
\renewcommand{\=}[1]{\stackrel{#1}{=}}

\newcommand{\didv}{\ensuremath{\mathrm{d}I/\mathrm{d}V}\xspace}

\makeatother

\usepackage{babel}

\newcommand{\beginsupplement}{%
	\setcounter{table}{0}
	\renewcommand{\thetable}{S\arabic{table}}%
	\setcounter{figure}{0}
	\renewcommand{\thefigure}{S\arabic{figure}}%
	\setcounter{equation}{0}
	\renewcommand{\theequation}{S\arabic{equation}}%
	\setcounter{section}{0}
	\renewcommand{\thesection}{S\arabic{section}}%
}

\begin{document}

\title{From Shapiro steps to photon-assisted tunneling in microwave-driven atomic-scale Josephson junctions with a single (magnetic) adatom}

\author{Martina Trahms}
\email{martina.trahms@cea.fr}
\thanks{\\ Current address: Univ. Grenoble Alpes, CEA, Pheliqs/Lateqs, Grenoble, France  }
\affiliation{Fachbereich Physik, Freie Universit\"at Berlin, Arnimallee 14, 14195 Berlin, Germany}

\author{Bharti Mahendru}
\affiliation{Fachbereich Physik, Freie Universit\"at Berlin, Arnimallee 14, 14195 Berlin, Germany}

\author{Clemens B. Winkelmann}
\affiliation{Univ. Grenoble Alpes, CEA, Grenoble INP, IRIG-Pheliqs, Grenoble, France}

\author{Katharina J. Franke}
\affiliation{Fachbereich Physik and Halle--Berlin--Regensburg Cluster of Excellence CCE, Freie Universit\"at Berlin, Arnimallee 14, 14195 Berlin, Germany}

\date{\today}
\begin{abstract}
Ultra-small Josephson junctions are strongly influenced by noise and damping due to energy dissipation into the environment, which are expected to suppress phase coherence. Here, we investigate the coherence properties of atomic-scale Josephson junctions in a scanning tunneling microscope under microwave excitation. Plain Pb-Pb junctions exhibit hysteretic Shapiro steps as signature of a coherent resonant state. With increasing AC amplitude, phase coherence is reduced due to an increase of thermal fluctuations. In the presence of magnetic adatoms the Josephson coupling energy is reduced and quasi-particle tunneling is enhanced. With AC driving we observe a rapid suppression of coherence that we ascribe to photon-assisted quasi-particle tunneling through Yu-Shiba-Rusinov states. Our results highlight the presence of phase coherence and shed light on the origin of the transition to incoherent transport, thereby revealing the importance of controlling dissipation in nanoscale superconducting devices.

\end{abstract}

\maketitle

Josephson junctions consist of two weakly coupled superconductors \cite{Josephson1962}. They enable tunneling of Cooper pairs and support a supercurrent that flows without dissipation due to a constant phase difference across the junction. Their functionality as building blocks in superconducting electronics and quantum technologies relies critically on maintaining the phase coherence. However, the phase coherence and superconducting transport in ultra-small Josephson junctions is crucially affected by the interaction with the environment, such as noise and dissipation \cite{McCumber1968, Devoret1990, Ingold1991, Grabert1999, Chauvin2006,Kautz1990}. 

In particular, atomic-scale Josephson junctions in a scanning tunneling microscope (STM) have a small capacitance, and thus a large charging energy. They strongly interact with the electromagnetic environment leading to energy dissipation and quantum effects \cite{Naaman2001, Ast2016, Jaeck2017}.
However, these STM junctions provide a large flexibility owing to their widely tunable conductance over several orders of magnitude and the capability to atomically engineer the junctions by including, for instance, magnetic atoms \cite{Randeria2016, Kuester2021, Karan2022, Trahms2023}. Magnetic adatoms introduce Yu-Shiba-Rusinov (YSR) states inside the superconducting junction \cite{Yu1965, Shiba1968, Rusinov1969}, which support additional channels of quasi-particle tunneling via resonant Andreev reflections \cite{Randeria2016, Ruby2015}. As the YSR states are also particularly attractive for engineering diode-like behavior in Josephson junctions \cite{Trahms2023, Steiner2023}, their impact on phase coherence is of special interest. 

From a fundamental perspective, the loss of phase coherence serves as an indicator of dissipation and noise, providing insights into the underlying microscopic transport mechanisms. In the transition from coherent to incoherent superconducting transport, the dissipationless supercurrent is gradually replaced by a resistive behavior originating from phase diffusion. This transition can be probed by high-frequency (HF) driving at angular frequency $\omega$. Indeed, flat Shapiro plateaus at multiples of voltage $V_n=\hbar\omega/(2e)$ in the voltage-current characteristics are the hallmark of phase coherent Cooper-pair tunneling \cite{Shapiro1963}. On the other hand, incoherent absorption of photons leads to resonances at the same voltage values, the effect being termed photon-assisted tunneling \cite{Tien1963, Falci1991, Roychowdhury2015, Kot2020, Peters2020, Siebrecht2023}. 

Here, we investigate the phase dynamics of current-biased Josephson junctions in an STM with and without magnetic adatoms in the presence of HF irradiation. We observe Shapiro steps and resolve the transition to incoherent transport at elevated AC driving amplitude. In the presence of a magnetic atom in the junction, the Josephson energy is reduced and, therefore, incoherent processes in the form of photon-assisted tunneling dominate the phase dynamics earlier.

\begin{figure}
\includegraphics[scale = 0.9]{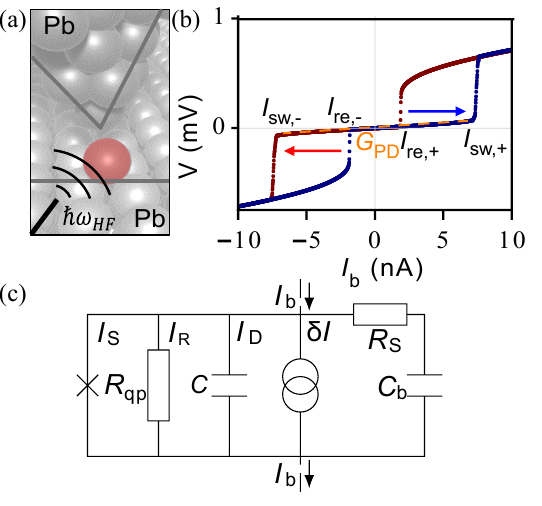}
\caption[Josephson Setup schematic]{Experimental realization of an AC driven Josephson junction in STM: a) Josephson junctions are formed between a Pb STM tip and a Pb(111) surface with an additional atom for stabilizing the junction. An antenna close to the junction is used to irradiate the junction with frequencies between 30\,GHz and 40\,GHz. b) Representative $V(I)$ curve  with $I_\mathrm{sw}$ and $I_\mathrm{re}$ currents marked in both biasing ($+/-$) direction. A finite slope around zero shows the presence of phase diffusion in the trapped state. This curve was recorded at 60\,$\mu$S (feedback opened at 10mV and 600nA). To capture the statistics the presented data shows the average of 100 sweeps in each direction. c) Equivalent circuit of the Josephson junction with an additional $R_\mathrm{S}-C_\mathrm{b}$ component in parallel to the common RCSJ model to account for frequency-dependent damping. }
\label{fig:JJSetup}
\end{figure}

We create atomic-scale Josephson junctions in an STM at 1.3\,K by bringing a superconducting Pb-coated tip, which has been prepared through repeated indentations into a Pb(111) substrate, into contact with the same substrate decorated with a single Pb atom. A key signature of a voltage-biased Josephson junction is the appearance of a zero-bias conductance peak. According to the Josephson relation $V_\mathrm{b}=\hbar \dot\varphi/2e$, applying a voltage bias induces a time-dependent phase change ($\dot\varphi$). As a result, voltage-biased measurements are not appropriate for probing phase dynamics or coherence in Josephson junctions. 
Instead, we investigate current-biased junctions, which we obtain by inserting a large ohmic resistor (1\,M$\Omega$) in the bias line. We then measure the resulting voltage drop across the junction using a differential amplifier. The voltage drop indicates that the phase across the junction is changing over time.

A representative $V(I)$ curve from an average of 100 spectra taken after approaching the tip above the superconducting gap at a bias voltage $V_\mathrm{b} = 10\,mV$ and current $I_\mathrm{set}$= 600\,nA (leading to a normal state conductance of 60\,$\mu$S) is shown in Fig.\,\ref{fig:JJSetup}b. Upon increasing the biasing current, a small voltage appears across the junction until a sudden rise marks the transition from the superconducting to the resistive (sub gap) state. The corresponding current is referred to as switching current ($I_\mathrm{sw}$). We note that $I_\mathrm{sw}\approx 7.3$ nA is much smaller than the critical current ($I_c \approx 127$\,nA) estimated from the Ambegaokar-Baratoff formula \cite{Ambegaokar1963} because of premature switching due to (thermal) noise. As the bias current is reduced, the junction remains in the resistive state until the voltage sharply decreases, indicating retrapping into the superconducting state. Since the retrapping current ($I_\mathrm{re}$) is smaller than the switching current, the junction exhibits pronounced hysteresis. This behavior is found regardless of the biasing direction of the current. We note that a small voltage persists even in the superconducting state, indicating the presence of a finite dissipative conductance ($G_\mathrm{PD}$) even in the superconducting state. 

The overall behavior of the junction can be understood considering the equivalent circuit diagram in Fig.\,\ref{fig:JJSetup}c with an ohmic resistor ($R_\mathrm{qp}$) representing the quasi-particle current in parallel to the supercurrent $I_\mathrm{S}$, and a capacitor $C$ reflecting the stray capacitance of the junction leading to a displacement current $I_\mathrm{D}$. Additionally, current fluctuations $\delta I$ due to noise and finite temperature are present in the experiment.
The simultaneous observation of a finite voltage associated to phase diffusion (finite $\dot\varphi$) and hysteresis is evidence of frequency-dependent damping arising from the junction's interaction with its electromagnetic environment. This interaction is modeled in the equivalent circuit by an $R_\mathrm{S}-C_\mathrm{b}$ filtering path, which leads to overdamped behavior at high frequencies and underdamped behavior at low frequencies \cite{Kautz1990}. 

\begin{figure*}
\includegraphics[width=0.9\textwidth]{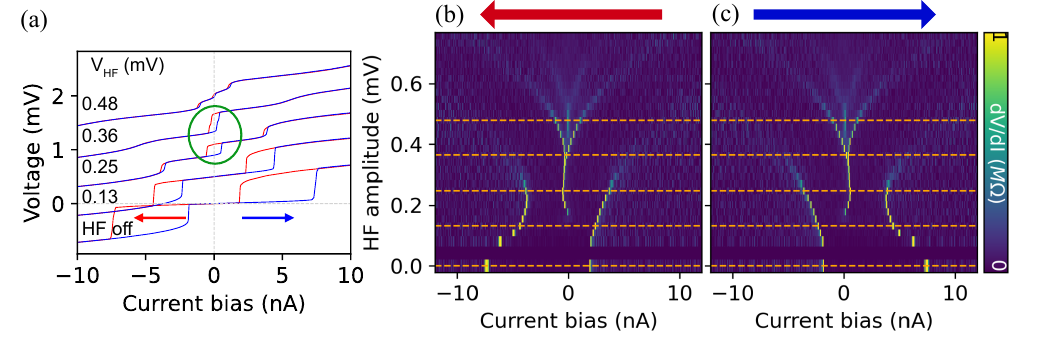}
\caption{Response of Pb-Pb Josephson junction to AC driving at 40\,GHz. a) Selected $V(I)$ curves subject to different amplitudes of AC driving. With increasing AC amplitude, $I_\mathrm{sw}$ is reduced and $I_\mathrm{re}$ increased. Hysteretic transitions between Shapiro steps can be observed at 0.25\,mV with vanishing hysteresis at larger AC amplitude. Curves are offset by 0.5\,mV for clarity, hysteretic Shapiro transitions are marked by a green circle. 
Upward sweeps are marked in blue while downward sweeps are shown in red. The bottom curve shows the same data as shown in Fig \ref{fig:JJSetup}b and all curves are recorded at a normal state conductance of 60\,$\mu$S and averaged over 100 sweeps.
b,c) 2D-colormaps of the numerical derivative $\mathrm{d}V/\mathrm{d}I$ of the recorded $V(I)$ curves with increasing AC amplitude of the two sweep directions indicated by colored arrows. The extracted curves in (a) are highlighted by orange lines. All spectra recorded at 60\,$\mu$S normal state conductance (at $V_{\text{bias}}$= 10\,mV).
}
\label{fig:HF_maps_Pb}
\end{figure*}

The phase dynamics of such Josephson junctions is well described by a modified resistively- and capacitively-shunted Josephson (RCSJ) model \cite{Stewart1969, McCumber1968, Ambegaokar1969, Ivanchenko1969, Kautz1990}. 
\begin{equation}
    (\hbar C/2e)\ddot \varphi + (\hbar /2eR_\mathrm{qp})\dot \varphi + I_\mathrm{S}(\varphi) +\delta I = I_\mathrm{b}.
    \label{Eq:RCSJ}
\end{equation}
Solving the RCSJ equation yields the dynamics of the superconducting phase ($\varphi$). The solution can be visualized using the mechanical analogy of a phase particle moving in a washboard potential, where the dynamics of the superconducting phase resemble those of a particle in a tilted, periodic potential with the depth of the potential minima given by the Josephson energy ($E_\mathrm{J}$). The potential is tilted by applying the biasing current ($I_\mathrm{b}$). Ideally, at small tilting, the phase particle remains trapped in one of the potential minima. Hence, the phase remains constant, resulting in zero voltage across the junction and dissipationless supercurrent flow. Strongly tilting the potential by a large current bias leads to vanishing potential barriers between the minima. The phase particle is then running down the tilted potential leading to a large voltage, and, hence dissipative current.

Interestingly, in our junction, we observe a finite voltage even in the superconducting regime. This suggests that the phase is not strictly pinned but instead undergoes diffusion between neighboring minima, driven by thermal fluctuations and noise before being retrapped again due to damping as a result of energy dissipation into the environment. The finite phase diffusion conductance signifies that phase coherence is perturbed by noise but not completely destroyed as in the freely running normal state. 

To probe and influence the phase coherence of the junction, we expose it to HF radiation of 40\,GHz via an antenna close to the STM junction \cite{Peters2020}. We then probe the response upon increasing the AC amplitude. (For a calibration of the AC amplitude, see SM) A few selected averaged $V(I)$ curves are shown in Fig.\,\ref{fig:HF_maps_Pb}a, while sets with continuously increasing AC amplitude are shown as color-coded maps of the derivative ($\mathrm{d}V/\mathrm{d}I$) curves in Fig.\,\ref{fig:HF_maps_Pb}b,c for the two ramping directions of the current bias, respectively. In the presence of weak HF radiation, the switching current is reduced whereas the retrapping current is increased (see, e.g., curve at 0.13\,mV AC amplitude). At the same time the slope in the superconducting regimes increases, indicating a larger phase diffusion. 

In the washboard-potential picture, periodically changing the periodic barrier height with the given amplitude and frequency, enhances both the probability of the phase particle to leave its local minimum and to be retrapped due to the periodic change in effective potential barrier height. This does not only affect the threshold currents but also increases the phase diffusion.  

Further increasing the AC amplitude enhances these trends, eventually leading to a vanishing hysteresis with overlapping switching and retrapping currents.
Once switching and retrapping currents occur at the same current-bias values, they start to shift towards larger current values with increasing AC amplitude. This suggests the dominance of the retrapping current when premature switching events are essentially possible at all times \cite{Iansiti1989}.
The steps eventually broaden until they are not discernible anymore from the overall increased slope due to strong phase diffusion. 

We note that the HF radiation not only modulates the potential at the AC frequency, but that it may also lead to heating of the junction, i.e., to random fluctuations. Both effects would lead to increased phase diffusion, reduced switching currents and increased retrapping currents as observed in Fig\,\ref{fig:HF_maps_Pb}a. 

\begin{figure*}
\includegraphics[width=0.9\textwidth]{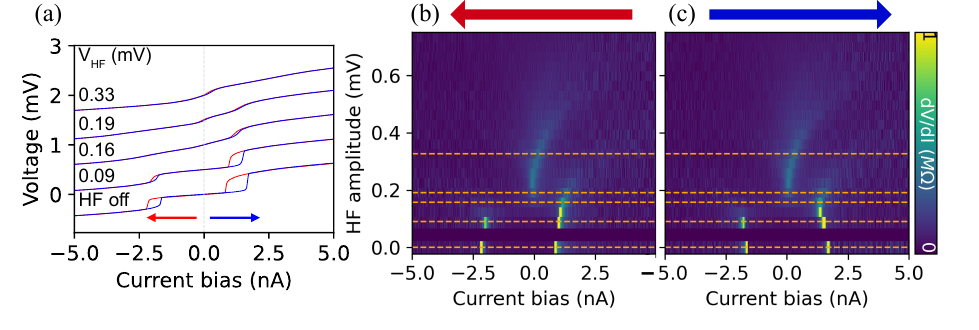}
\caption[2D-colormaps of a current-biased Pb-Mn-Pb Josephson junction under HF irradiation]{Response of Pb-Mn-Pb Josephson junction to AC driving at 40\,GHz. a) Selected $I(V)$ curves for different AC amplitude. Hysteretic transitions between Shapiro steps can be observed. Curves are offset by 0.5\,mV for clarity. All curves are recorded at a normal state conductance of 50\,$\mu$S and averaged over 100 sweeps.
Upward sweeps are marked by blue while downward sweeps are shown in red. b,c) 2D-color maps of $\mathrm{d}V/\mathrm{d}I$ curves with increasing AC amplitude at the two sweep directions indicated by colored arrows. The extracted curves in (a) are highlighted by orange lines in (b) and (c).
All junctions are current-biased at 50\,$\mu$S normal state conductance (at $V_{\text{bias}}$= 10\,mV).
}
\label{fig:HF_maps_Mn}
\end{figure*}

While the so-far observed changes could, thus, be a signature of a loss of phase coherence, we also find indications of preserved coherence that we explain in the following. With increasing AC amplitude, we observe additional steps and corresponding plateaus at low biasing current in the $V(I)$ curves with a hysteresis opening around zero bias. This implies that the junction remains in the state of the lowest plateau when decreasing the bias and even when reversing the biasing direction.
As explained above, the HF radiation periodically modulates the slope of the washboard potential. We, thus, need to add the corresponding AC driving current $I_\mathrm{AC} \sin{\omega_\mathrm{AC}t}$ current to equation (1). 
When the average voltage from the motion of the phase matches  $V_n= n\frac{\hbar\omega}{2e}$, a resonance condition is reached, where the motion of the phase particle is synchronized with the AC drive. At these points, steps in the $V(I)$ mark the transitions between discrete voltage resonances at $V_n= n\frac{\hbar\omega}{2e}$. These steps are known as Shapiro steps and are a signature of coherent phase dynamics \cite{Shapiro1963}.
The coexistence of hysteresis and phase diffusion on the Shapiro plateaus is again evidence of the junction experiencing frequency-dependent damping \cite{Kautz1990, Hamilton2000, Larson2020}. Hence, although we have signatures of phase coherence, the latter is gradually reduced with increasing AC amplitude reflected by the increasing slope of the plateaus \cite{Koval2004}. 

As the AC amplitude increases further the hysteresis associated to Shapiro plateaus, as highlighted in Fig.\,\ref{fig:HF_maps_Pb}a, gradually disappears. This goes along with a transition to completely incoherent photon-assisted tunneling processes. The loss of coherence may be enhanced by heating and associated thermal noise.

As the next step, we aim to investigate how the transition from coherent to incoherent phase dynamics is affected by the presence of a magnetic adatom in the junction. To start with, the presence of an unpaired spin reduces the Josephson energy \cite{Glazman1989,Choi2004,vanDam2006,Cleuziou2006,Jorgensen2007}, and thus the corrugation of the washboard potential. Further, it induces YSR states within the superconducting  gap. The quasi-particle tunneling probabilities involving a YSR state generally depend on the tunneling direction \cite{Ruby2015}, leading to an asymmetric shunt conductance in the RCSJ model and, thus, non-reciprocal damping in the Josephson junction. This asymmetry manifests for instance as distinct retrapping currents depending on the direction of the applied current \cite{Trahms2023, Steiner2023}. 

To investigate this, we perform the same measurements on junctions, with a single Mn atom with nearly identical normal-state tunneling conductances (50 $\mu$S, instead of 60 $\mu$S previously).  
As expected, we note a significant reduction in the switching current for the Mn junction (Fig.\,\ref{fig:HF_maps_Mn} a bottom curve; for a direct comparison to Pb-Pb see SM), due to the reduction of the Josephson coupling energy \cite{Randeria2016, Karan2022}. This implies weaker phase locking, and thus stronger phase diffusion. Furthermore, the aforementioned non-reciprocal damping behavior manifests in the asymmetric retrapping currents \cite{Trahms2023}. 

Under HF driving, the switching current is further reduced, in a similar way to the non-magnetic case (Fig.\,\ref{fig:HF_maps_Mn}a). However, the increase in phase diffusion is less pronounced, probably because of its higher initial level in the Mn junction. Upon further increasing the AC amplitude, non-linearities appear close to zero bias (Fig.\,\ref{fig:HF_maps_Mn} 0.16\,mV and 0.19\,mV AC amplitude), but are too smeared to form plateaus. This is particularly evident from the $\mathrm{d}V/\mathrm{d}I$ curves in Fig.\,\ref{fig:HF_maps_Mn}b,c, where only one broad peak appears close to zero bias and further broadens and shifts with increasing amplitude. The maps also clearly bring out the asymmetric nature of the junction. At negative bias the hysteretic switching between superconducting and normal state is quickly merging into the incoherent background at AC driving leading to the absence of a clear feature, in contrast to features at positive voltages. 

To understand this behavior, we consider the RCSJ model including a non-ohmic resistor $R_{\rm qp}$ associated to quasi-particle tunneling (Fig.\,\ref{fig:JJSetup}c). Under AC driving, $R_{\rm qp}$ is modified according to the Tien-Gordon model \cite{Tien1963, Falci1991}, where photon-assisted tunneling leads to a splitting of the coherence peaks and subgap conduction resonances originating from the YSR states or multiple Andreev resonances. In voltage-biased measurements, this splitting can be directly observed \cite{Peters2020, Siebrecht2023} (see also SM). In the presence of the Mn adatoms with asymmetric YSR intensities, the Tien-Gordon pattern is also strongly asymmetric due to resonant Andreev reflections \cite{Ruby2015, Peters2020}. 
As a result, photon-assisted quasi-particle tunneling through YSR states becomes the dominant transport channel, smearing out the $V(I)$ characteristics and preventing phase coherence.

In conclusion, we have probed the response of atomic-scale Josephson junctions formed with a superconducting Pb tip on a Pb(111) substrate decorated with individual magnetic and non-magnetic atoms. Despite of thermal fluctuations at 1.3\,K and strong coupling to the electromagnetic environment, we observe Shapiro steps with hysteretic plateaus for low amplitudes of AC driving indicating finite superconducting phase coherence in these junctions. Yet, we also show that phase coherence is fragile and strongly affected by thermal fluctuations at higher HF amplitudes. 

In Mn-decorated junctions, reduced switching currents and enhanced phase diffusion point to a smaller Josephson coupling and shorter phase coherence. Upon applying HF radiation, we observe a more rapid poisoning of coherence leading to smeared, asymmetric $V(I)$ curves. We ascribe their asymmetry to photon-assisted quasi-particle tunneling through YSR states, which dominate transport as the AC drive splits the in-gap states. 
These findings demonstrate how coherent Josephson dynamics give way to incoherent, photon-assisted transport under the influence of YSR states and environmental effects, emphasizing the need for careful engineering of the junction environment to preserve phase coherence in atomic-scale superconducting devices.

\section*{Acknowledgments}
We thank S. Cailleux, J. I. Pascual, S. Trivini and W. van Weerdenburg for discussions, and C. Lotze for general technical support. We acknowledge financial support by the Deutsche Forschungsgemeinschaft (DFG, German Research Foundation) through projects FR2726/11-1, as well as by the joint project of DFG-Agence Nationale de la Recherche under grant FR2726/10-1 and ANR-17-CE30-0030, respectively.

%

\clearpage

\onecolumngrid

\newcommand{\vsigma}{\mbox{\boldmath $\sigma$}}

\beginsupplement
\section*{{Supplemental Material}}

\maketitle 

\section{Experimental details}
The Pb(111) crystal was cleaned by repeated cycles of sputtering with Ne$^+$ ions and annealing at 400\,K. The sample was then transferred under ultra-high vacuum into the STM, which is operated at a base temperature of $T= 1.3\,\mathrm{K}$. At this temperature, the superconducting energy gap of Pb is $\Delta = 1.35\,\mathrm{meV}$. 

Superconducting Pb tips were prepared by indenting the tip wire into the Pb(111) substrate at an applied voltage of 100\,V. The procedure was repeated until the gap of the tip was the same as of bulk Pb.

Pb atoms were dropped from the tip when approaching the substrate with high current values. Mn adatoms were deposited on the clean surface with the sample taken out of the STM and placed on a liquid-Helium cooled manipulator. Two different adsorption sites of Mn atoms on Pb(111) exist with different apparent heights and different spectroscopic fingerprints. The sites can be manipulated by approaching the Mn atom with the STM tip. The Josephson measurements were done on the higher adsorption configuration as it is the only stable one under the conditions for forming a Josephson junction \cite{SRuby2015}.

The high-frequency radiation was applied by an external signal generator (R\&S\textsuperscript{\textregistered}SMB100A) through high-frequency-optimized cables. The open end of a superconducting Nb-Ti coaxial cable close to the STM junction acts as an antenna \cite{SPeters2020}.

\section{Current-biased measurements}

\begin{figure}[h]
\includegraphics[scale=0.7]{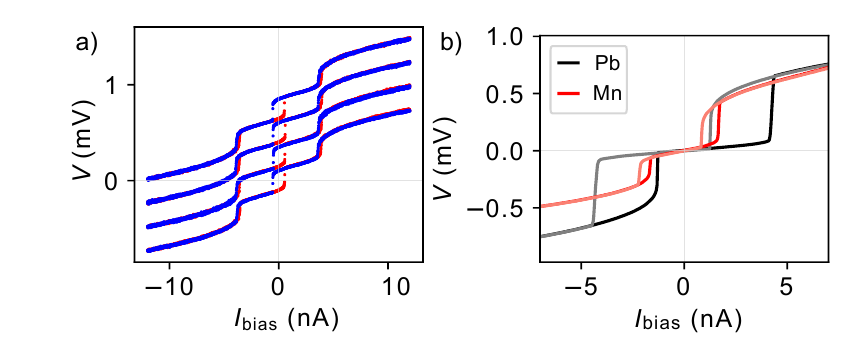}
\caption{a) $V(I)$ characteristics of single (not averaged) current-bias sweeps on the 60\,$\mu$S Pb junction in the presence of finite HF irradiation ($V_\mathrm{HF}$ = 0.25\,mV). The sweeps are offset by 0.5\,mV for better visibility. b) Comparison of Josephson junctions formed on a Pb and Mn adatom for both sweep directions at 50\,$\mu$S. The presence of the magnetic adatom reduces $I_\mathrm{sw}$ and increases the phase diffusion slope.}
\label{fig:S1}
\end{figure}

If not indicated otherwise in the text, Josephson junctions were formed by setting the desired normal state conductance at a bias voltage of 10\,mV, which is well outside the superconducting gap of Pb.
After waiting a few minutes to stabilize the current at the desired junction conductance, the feedback is turned off, and the 1\,M$\Omega$ resistor is inserted in the bias line for current-biased measurements.
While sweeping the effective current the voltage in the junction is measured by a differential amplifier. As switching and retrapping events are of statistical nature, at least a hundred repetitions of $V(I)$ curves were collected in the same junction. 
After that the junction is set back to voltage-bias mode by by-passing the 1\,M$\Omega$ resistor and activating the feedback loop.

Single $V(I)$ curves (i.e. not averaged) of Pb--Pb junctions in the presence of HF radiation of 0.25\,mV amplitude are shown in Fig.\,\ref{fig:S1}a. In the main text, we show the average of 100 such sweeps (Fig.\,2a and Fig.\,3a).

Fig.\,\ref{fig:S1}b compares a Pb--Pb junction with one including a Mn atom without HF radiation (average of 100 sweeps) at a normal state conductance of 50\,$\mu$S (10\, mV, 500\, nA). Both spectra show hysteretic behavior in the switching/retrapping current. Switching currents are significantly suppressed and phase diffusion is enhanced in the presence of the magnetic adatom. This indicates a reduction of the Josephson coupling energy in the presence of the unpaired electrons of the Mn atom as explained in the main text. While the Pb--Pb junction is symmetric in biasing direction, the Pb--Mn--Pb junction is non reciprocal due to quasi-particle tunneling through YSR states and associated damping \cite{STrahms2023}.

We note that the current bias that is applied is not perfect, in particular outside of the trapped state when the resistance of the STM junction becomes non-negligible with respect to the biasing resistor. 
This results in a systematic error on the resistive parts of the junction but does not change the experimental finding of this work.

\section{Calibration of AC amplitude}
\begin{figure}[h]
    \includegraphics[scale =0.8]{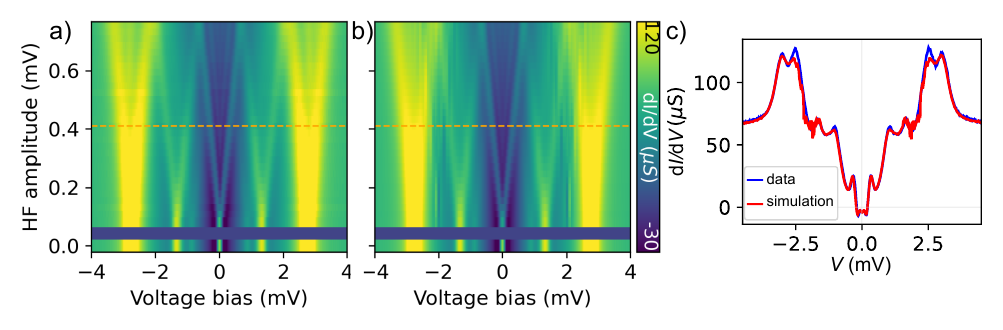}
    \caption{Measured d$I$/d$V$ spectra in the presence of HF irradiation in comparison to the calculated spectra using the Tien-Gordon model. a) shows the measured data of a Pb-Pb junction at 60\,$\mu$S under 40\,GHz HF irradiation. In b) the Tien-Gordon model was used to calculate the photon-assisted tunneling processes for the same junction starting from the experimental d$I$/d$V$ spectrum at zero HF amplitude. For Andreev and Cooper-pair tunneling processes the transmission of two electrons was assumed in the calculation in b). In c) an exemplary data set at finite amplitude (red dashed line in a and b) comparing the measured data to the simulated data is shown. This comparison was used to determine the damping properties of the HF cabling in the cryostat. }
    \label{fig:S2}
    \end{figure}

To calibrate the amplitude of the HF radiation we measured voltage-biased \didv spectra with increasing power of the signal generator. One set of spectra on a Pb--Pb Josephson junction at 60\,$\mu$ S normal-state conductance is shown in Fig.\,\ref{fig:S2}a. They exhibit the typical V shaped splitting of the quasi-particle, Josephson and Andreev peaks with increasing HF amplitude. The pattern is simulated by the Tien-Gordon model from the experimental data in the absence of HF irradiation at that junction conductance. The HF amplitude determines the size of the splitting. The precise pattern of the splitting depends on the effective charge of the transported quasi particle, i.e., 2e for the Josephson peak and even higher charges for higher-order Andreev reflections. Since the most dominant peaks in our spectra are the Josephson peak and the first Andreev reflection, a charge of 2e has been taken into account for the in-gap features in the simulation shown in Fig.\,\ref{fig:S2}b. The simulation procedure to determine the HF amplitude is described in more detail in \cite{SPeters2020}.  An exemplary spectrum and the corresponding Tien-Gordon fit is shown in Fig.\,\ref{fig:S2}c. The fit yields the effective HF amplitude in the junction. From this the damping of the cables all the way from the signal generator through the cryostat to the antenna and the coupling efficiency to the junction can be determined. 

\section{YSR states of Mn atoms on Pb(111)}
\begin{figure}[h]
    \includegraphics[scale =0.8]{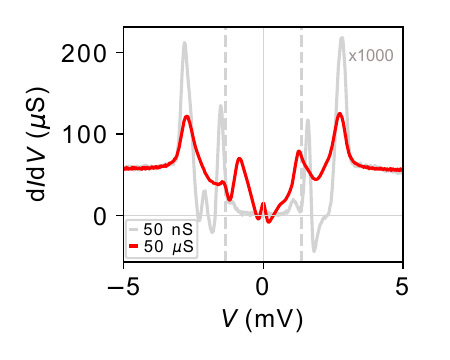}
    \caption{d$I$/d$V$ spectra at low (50\, nS) and high (50\,$\mu$ S) junction conductance on a single Mn adatom without HF irradiation. In the far tunneling regime, the YSR states are reflected in conductance peaks at symmetric bias voltage but asymmetric intensity (grey). Resonant Andreev reflections through the YSR states lead to an asymmetric spectrum also at high junction conductance, i.e., in the Josephson regime.}
    \label{fig:S3}
    \end{figure}

Exchange coupling of magnetic adatoms induce YSR states inside the superconducting energy gap of the Pb substrate. These are well resolved in voltage-biased junctions in the tunneling regime (Fig.\,\ref{fig:S3}, grey). As the tip--sample distance is decreased additional Andreev related tunneling processes occur leading to an enhanced conductance at energies well below the tip gap. Resonant Andreev reflections through the electron-hole asymmetric YSR states and their thermal replica imprint a characteristic asymmetric line shape in the spectra. 

\section{Photon-assisted tunneling through YSR states in the Josephson regime}

\begin{figure}[h]
    \includegraphics[scale =1]{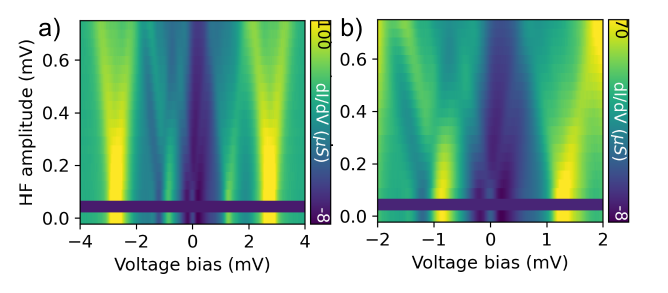}
    \caption{\didv spectra recorded on Mn adatoms in a junction of 50\,$\mu$S with 40\,GHz irradiation. b) Close-up view on the smaller voltage values. Photon-assisted tunneling splits all tunneling paths. Due to the low-lying resonant Andreev reflections, the photon-assisted tunneling processes overlap with the splitting of the Josephson peak.}
    \label{fig:S4}
    \end{figure}

Fig.\,\ref{fig:S4} shows voltage-biased \didv spectra of a Pb--Mn--Pb junctions in the presence of 40\, GHz irradiation at 50\,$\mu$S. All resonances are split due to photon-assisted tunneling according to Tien-Gordon theory as the HF amplitude is increased. Because the YSR states cause resonances that are quite close to zero bias, intensities start overlapping already at low HF amplitudes as emphasized by the zoom in on Fig.\,\ref{fig:S4}b.

\end{document}